\documentclass{article}
\usepackage{graphicx}
\usepackage{amsthm}
\usepackage{amsfonts}
\usepackage{amsmath}
\usepackage{amssymb}
\usepackage{fullpage}
\usepackage[usenames]{color}
\usepackage{multirow}
\usepackage{hyperref}
\usepackage{subcaption}
\usepackage{setspace}

\hypersetup{
	colorlinks = true,
	urlcolor = blue,       
	linkcolor= blue,       
	citecolor= blue,       
	filecolor= blue,        
}

\usepackage{listings}

\definecolor{dkgreen}{rgb}{0,0.6,0}
\definecolor{gray}{rgb}{0.5,0.5,0.5}
\definecolor{mauve}{rgb}{0.58,0,0.82}

\title{Artificial Intelligence for Automatic Detection and Classification Disease on the X-Ray Images.}        
\author{Liora Mayats-Alpay\\ 
\\Fowler School of Engineering.
Computer Science.
    \\Chapman University
    \\mayatsalpay@chapman.edu}

\date{}

\begin{document}
	
	\maketitle
	
	
	\doublespacing
	\begin{abstract}

Detecting and classifying diseases using X-ray images is one of the more challenging core tasks in the medical and research world. Due to the recent high interest in radiological images and AI, early detection of diseases in X-ray images has become notably more essential to prevent further spreading and flatten the curve. Innovations and revolutions of Computer Vision with Deep learning methods offer great promise for fast and accurate diagnosis of screening and detection from chest X-ray images (CXR). This work presents rapid detection of diseases in the lung using the efficient Deep learning pre-trained RepVGG algorithm for deep feature extraction and classification. We used X-ray images as an example to show the model's efficiency. To perform this task, we classify X-Ray images into Covid-19, Pneumonia, and Normal X-Ray images. For evaluation, first, we used a histogram-oriented gradient (HOG) to detect the shape of the region of interest (ROI). Employ ROI object to improve the detection accuracy for lung extraction, followed by data pre-processing and augmentation. Then, a pre-trained RepVGG model is used for deep feature extraction and classification, similar to VGG and ResNet convolutional neural network for the training-time and inference-time architecture transformed from the multi to the flat mode by a structural re-parameterization technique. Next, we created a feature map using the Computer Vision technique and superimposed it on the original images. We are applying Artificial Intelligence technology for automatic highlighted detection of affected areas of people's lungs. Based on the X-Ray images, an algorithm was developed that classifies X-Ray images with height accuracy and power faster thanks to the architecture transformation of the model. We compared deep learning frameworks' accuracy and detection of disease.
The study shows the high power of deep learning methods for X-ray images based on COVID-19 detection utilizing chest X-rays. The proposed framework offers better diagnostic accuracy by comparing popular deep learning models, i.e., VGG, ResNet50, inceptionV3, DenseNet, and InceptionResnetV2.

	\end{abstract}
	
	\bigskip


	
	
	\section{Introduction}\label{intro}


The medical diagnosis of X-Ray images plays a significant role in the health care system. Rapidly growing technologies and the ability to innovate with AI and  Machine learning tools make medical  process X-Ray images faster and more efficient. The aims of AI in the medical field are the early and fast diagnosis of illnesses, which greatly benefits doctors and patients. Driving innovation allows scientists to create deep learning algorithms that automatically classify and detect diseases from big data X-Ray images.

The main benefits of innovation in the medical and engineering field are saving time, cost, and life of people. The core of deep learning was taken from the biological system of neurons in our brain, which takes information as input, performs a task between artificial neurons network (ANN) system, and get a result as output. The idea of AI is to create new deep-learning tools to perform tasks like humans. Many researchers used ANN to develop different methods to perform the task in X-ray images. The most popular network for image problems is Convolutional Neural Network(CNN). Convolutional Neural Networks (CNN) became very powerful in the imagery field and AI society; then, in the annual competition in 2014 AlexNet model won  ImageNet 2014 Challenge competition for image recognition. The AlexNet model evaluated the large-scale images using CNN architecture for image recognition task and is described in the paper Krizhevsky et al. \cite{Krizh}. Many scientists used AlexNet CNN model for challenging task image recognition. The paper of  Russakovsky et al. \cite{olga} uses the AlexNet model to evaluate and analyze the classification and detection object in the images. "We discuss the challenges of collecting large-scale ground truth annotation, highlight key breakthroughs in categorical object recognition, provide a detailed analysis of the current state of the field of large-scale image classification and object detection, and compare the state-of-the-art computer vision accuracy with human accuracy" \cite{olga}.
The base of CNN's architecture has greater benefits for the various deep learning image recognition problems, image classification and detection objects, image recognition, and image analysis in various fields, especially in the medical industry. In the work of He et al. \cite{Res}, the paper describes a deep learning framework for image recognition\cite{kaim}. The paper by He et al. applies different Machine learning techniques in Intelligent medicine to become powerful at deep frameworks and computer vision. They have become state-of-the-art for visual applications such as medical image analysis, image classification, and image recognition. The work of Simonyan, Karen, and Andrew Zisserman has explained the deep convolutional networks for large-scale image recognition \cite{Simon}. The consequence of the automatics detection process can work with an extensive database of patient X-Ray images. 

In the present work, we propose an algorithm that classifies and detect diseases in the X-ray images based on the deep learning RepVGG structure.
Using the public chest X-ray images dataset, we developed adaptive modules for the automatic detection of COVID-19 Pneumonia and Normal. Specifically, we automatically segment the lung region and locate the Covid-19 highlighted affected area, then construct a classification tool capable of assessing the presence of COVID-19, Pneumonia, and Healthy cases on scanning X-Ray images.

 To demonstrate how the algorithm performs the task was taken an example of the data of the X-Ray images of Covid-19, Pneumonia, and Normal cases.  The RepVGG architecture model was created by Din et al.\cite{Ding}. The specificity of this method is the ability to transform from multi branches to the plane architecture structure of the model. The structure was built from the idea of combining two models, VGG and ResNet. We will describe the architecture structure and functionality in the section 'RepVGG architecture transform from multi-branches to compact model.'

The present work proposes improving the lung domain detection diagnosis. Apply different computer vision techniques to detect affected areas of COVID-19, Pneumonia, and Normal cases in the lung. In computer vision and image processing, detecting the concept's target on the X-ray images could be challenging. We used the Histogram oriented gradient (HOG) for the Region of Interest (ROI) segmentation to detect the object lung region of interest extraction in the X-ray
images\cite{ha}\cite{ra}. The method a Gradient - Weighted Class Activation Mapping (Grad- Cam) described in the work of (\cite{selv}). Using this method, we created a heat map using a computer vision method to detect the affected area in the lung. We superimposed the heat map on the original chest X-ray image and visually explained it by highlighting the X-ray images to evaluate the proposed method. 
Examining radiological images requires a long experience to interpret an X-Ray (and in particular to distinguish between the case of Pneumonia and COVID-19), and not enough specialists are available, even in developed countries, let alone in the third world.  
Using Artificial Intelligence in medicine allows us to speed up the detection of COVID -19 on radiological images and work more efficiently. It can save lives and reduce time and cost. The consequence of such a rapid detection process can work with an extensive database of patient radiological images. Due to the importance of depth significance, much research focuses on recognizing Covid-19 and the classification of X-ray images using  Artificial Intelligence enhanced detection and various machine learning methods.
In this work, we propose some other configuration results of the computer visualization framework, what the machine learning predicts correct and wrong by the confusion matrix and classification report. 
Another direction of research in this paper is inspired by the work of (Ding. and al \cite{Ding} ) on the new architecture structure of the convolutional neural network RepVGG model\cite{ha}. The model achieves excellent accuracy performance, improves classification performance, and decreases the training time of the data set. This architecture structure is based on VGG and ResNet (CNN) models but is much more powerful with high-performance accuracy and smaller than the simple architecture structure\cite{res}. This algorithm uses branches and $3$x$3$ CNN and ReLU (Rectified Linear Unit). Different CNN models use multi-branches such as ResNet \cite{res}, Inception \cite{inc}\cite{inc4}, and DenseNet \cite{dence}, and we will compare their effectiveness and accuracy later. These architectures use multi-branch models with complex structures, extended computational time, require other operational processes, and reduce implementation tasks. 
The significant role of RepVGG is that it uses an additional $1$x$1$ CNN branch in architecture during the training process. After completing the learning process, the branches are deleted, and the architectural structure has reduced in size. Such transformation architecture structure allows the re-parameterization and optimization of the processes. This re-parameterization makes the method more efficient and faster in image classification and segmentation. The model becomes smaller and easier to use functionality by changing and reducing the size. As a result, this optimization process allows the signals in neural networks to be transmitted faster, saving time and reducing the cost for manufacturers and developers.
  \\



In this paper, we investigate X-Ray scans; using the Histogram oriented gradient (HOG) method, we show the Region of Interest (ROI) segmentation in X-Ray images. We then apply a Grad-Cam feature map to visualize and highlight the affected areas in the X-Rays. Using the RepVGG (a powerful architecture structure), we create an algorithm that classifies X-Ray images: Covid-19, Pneumonia, and Healthy cases. We show the ROC curve and AUC value for the proposed framework in classification and give a summary of the machine learning framework with a confusion matrix. Finlay, we present a table accuracy by comparing the deep learning models using multi-branches such as ResNet, Inception, and DenseNet-classification report with a comparison of different methods.

\section{Data and Methods}\label{data}



In our work, we proposed the automatic classification of X-Ray images using deep learning methods into three categories such as COVID-19, Pneumonia, and Healthy cases, by building an Artificial Neural Network model. Also, using Computer Vision techniques, we create an algorithm that detects lung diseases and visualizes the affected area on the X-ray images. The creation process consists of several stages, as shown in Figure 2. At the beginning of the process, we split the data into two subsets: train and test data. In the first step, we build the artificial training model using machine learning supervision methods with labeled images. In the second step, we used this creating model to test new data and predict the final output to classify and detect X-Ray images by three categories: Covid-19, Pneumonia, and Healthy cases.

Our work collected data from Kaggle\cite{kag} with a total number of 2856 X-Ray images used and illustrated in the Table in Figure 1. The data includes 1000 COVID-19 scanning x-ray images, 1000 normal images, and 856 pneumonia images. We split data into two subsets and used for the first subset 80\% of data for training and validation and the second subset 20\% of data for testing. The coding is carried out with the Pytorch package on a workstation equipped with the NVIDIA RTX A6000 with 512G RAM. We have trained the proposed model for 20 epochs and set the learning rate to 0.001. We propose visualization predict and show the graphs of the confusion matrix, ROC curves, and AUC values for the proposed framework in classifying COVID-19 X-Ray images. The achieved results we summarized in section Results and illustrated in Table I, Table II, and Table III of Figures 16, 17, and 18.  Performance comparison of the RepVGG model and other popular deep learning models in terms of precision, sensitivity, F1 score, and accuracy. The best results are labeled in bold. The proposed framework can classify all pneumonia and only misclassify one COVID-19 image. RepVGG shows the best performance in all metrics for COVID-19 and Pneumonia cases. Even for the normal cases, the model behaves better than others, with only the Precision being slightly lower than Inception ResnetV2.




    
    
    
    
    
    

\begin{figure}[!h]
\center
	\includegraphics[scale=0.3]{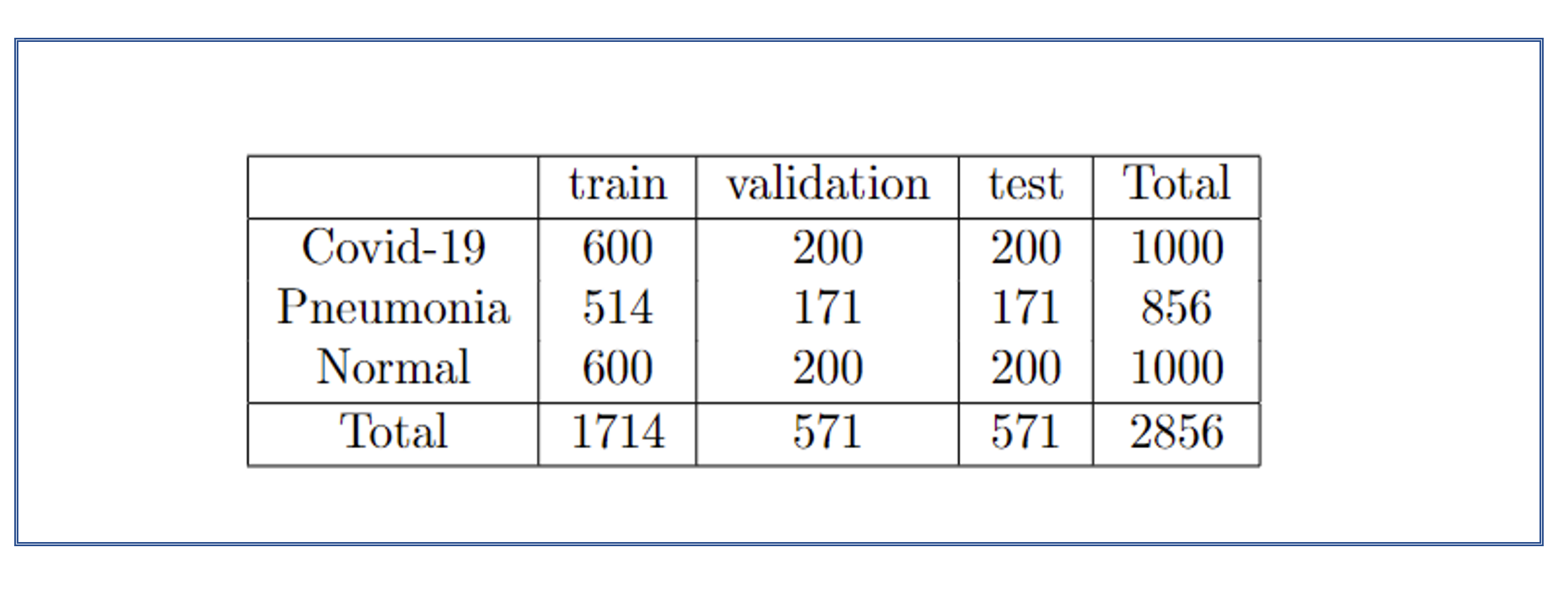}
	\caption{Table of data set.}
\end{figure}

\vfill

\subsection*{Methods}

In this paper, the proposed COVID-19 diagnosis prediction model consists of several steps, as shown in Figure 2. At the beginning of the process, we split the data set into two subsets: the first train and validation and the second test subset. The ratio of splitting data is training and validation subset 80\% of all data and the second test subset 20\%. We create the trained model using a supervised machine learning framework in the training validation subset. Then we build artificial neural network models consisting of steps such as image prepossessing, HOG lung area detection, and transfer learning.
Then we use the creating training model for the rest of the 20\% images in the test data set. In the test model, we use new data which newer training before and perform the same process as in the training model, such as image prepossessing, HOG lung area detection, and transfer learning.  In the test model, we predict the output classification of X-Ray images. We apply Convolutional Neural Networks to classify X-ray images in three categories: Covid-19, Pneumonia, and Normal (healthy). While creating the training model, we also used the deep learning computer visualization framework Card-Cam to visualize, detect and highlight the affected area in the lungs in the x-ray images.

 Then a pre-trained RepVGG model is applied and fine-tuned for COVID-19 diagnosis. RepVGG, recently developed by Ding et al\cite{Ding}, is a convolutional neural network with a VGG  plain model without any branches. The model’s inference-time body uses only $3 \times  3$ Conv and ReLU, while the training-time model has a multi-branch topology. Such decoupling of the training-time and inference-time architecture is realized by a structural re-parameterization technique, so the model is named RepVGG. The multi-branch is implemented by using identity and $1\times1$ branches. A core block compared with ResNet architecture is shown in Figure 3. This work reveals that deep learning shows great potential in detecting COVID-19 based on chest X-Ray images. The proposed framework shows the best diagnostic accuracy by comparing popular deep learning models, i.e., VGG, ResNet 50, inceptionV3, DenseNet, and Inception ResNetV2.

\begin{figure}[!h]
\center
	\includegraphics[scale=0.3]{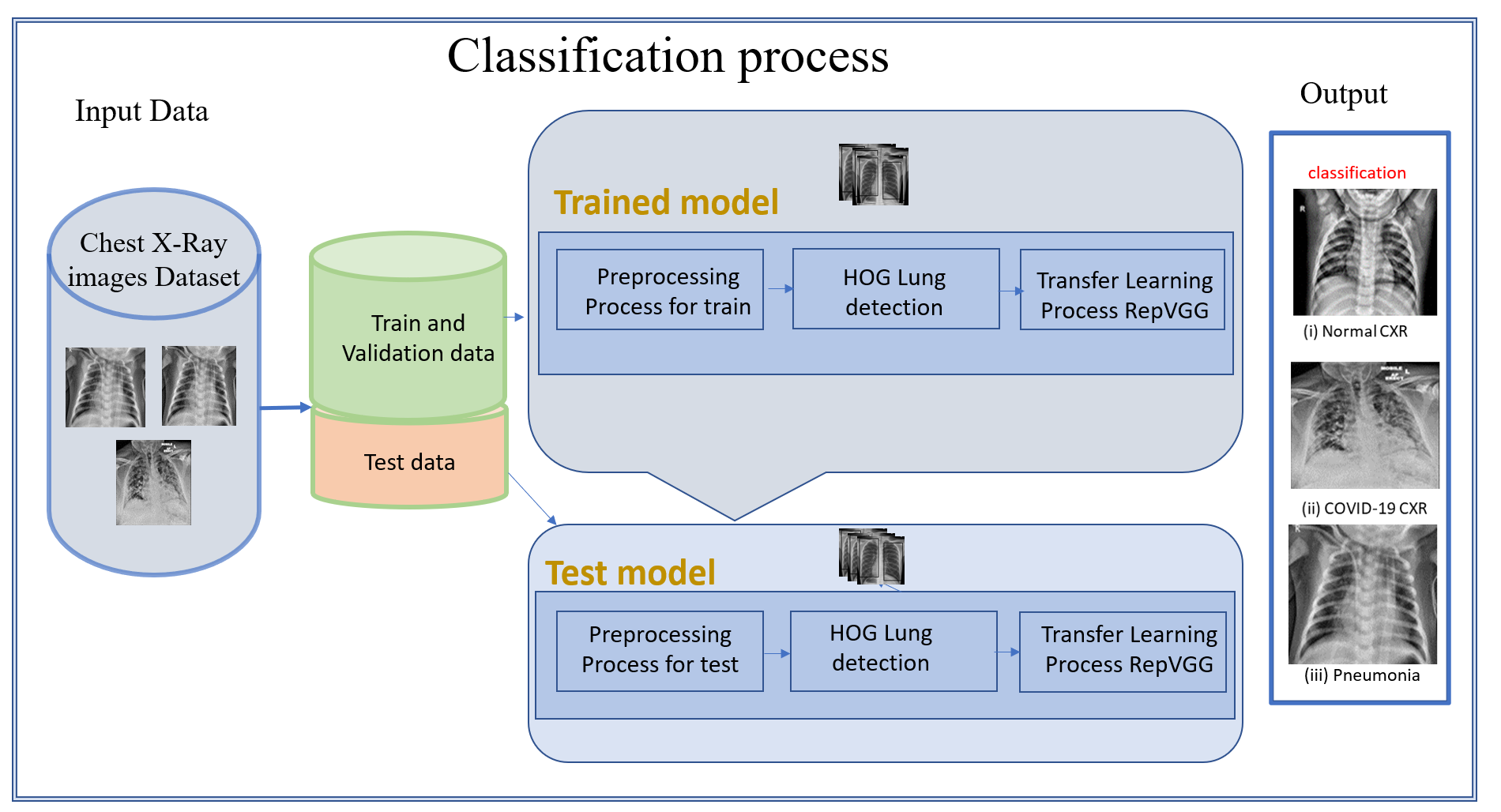}
	\caption{Classification process.}
\end{figure}

\subsection*{Image prepossessing process}


The image processing process aims to make the ability computer read the image from the source and feed it into the model. Usually, medical x-ray images come in grayscale. The image is divided by the pixels, and the color in the image is specified as an integer number of colors in the range between 0 and 255. The number 0 corresponds to the completely black color and 255 to the completely white color, and the number between these ranges corresponds to different shades of grey.
The prepossessing image task is to improve image features for the input image before building the model. 
Since each deep learning model has its special standard size according to which the model works. It is necessary to resize the image before feeding it into the respective model. The pre-processing process performs image classification by the transformations: resize, lightness, random rotation, shifting image position, flipping the image in horizontal or vertical space, and re-scaling the image data.
When performed correctly, the image pre-prepossessing process allows a decrease in the running time of the model and increases the efficiency of model classification.


\subsection*{Histogram-oriented gradient (HOG) for the Region of Interest (ROI)}


The Histogram oriented gradient (HOG) is a feature extraction method that is often used in computer vision and image processing for object detection tasks, such as describing the object shape by calculating different gradients. In the current work, we apply Region of Interest (ROI) extraction for the HOG to describe lung location in the X-ray images. One of the challenging tasks in X-Ray images is to find the two part of the lung as the object in the gray image and shape it. We applied several techniques to detect the shape of the lung in the scanning X-Ray images. This yields different results, shapes output, and illustrates a visualization of the HOG differently. 
Figure 3 in the left original X-Ray image, in the middle, resize the X-ray image and shape the 300 X 350 pixels and the right result as output X-Ray image. The same original X-Ray image in Figure 4 is shown in the middle, resized and shaped 100 X 150 pixels X-ray image, and in the right result as output X-Ray image. 
Figures 3 and 4 show that the same image processing HOG used a similar technique; depending on the resized X-ray image, we get different results.


 \begin{figure}[!h]
\centering
\begin{subfigure}[b]{0.2\linewidth}
 	\includegraphics[width=\linewidth]{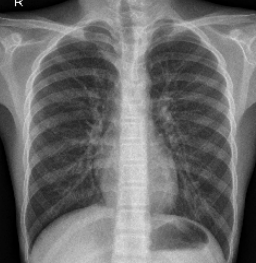}
	\caption{Original image}
\end{subfigure}
\hspace{5mm}
\begin{subfigure}[b]{0.2\linewidth}
	\includegraphics[width=\linewidth]{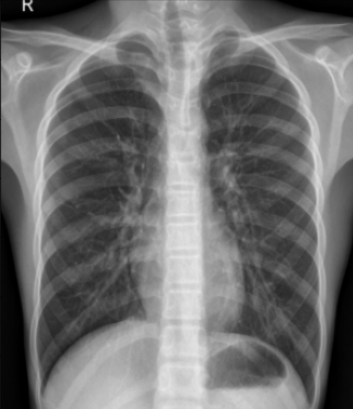}
	\caption{Resize image}
\end{subfigure}
\hspace{5mm}
\begin{subfigure}[b]{0.2\linewidth}
	\includegraphics[width=\linewidth]{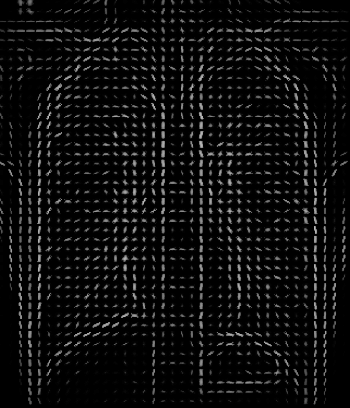}
	\caption{HOG}
	\end{subfigure}
	\caption{HOG  feature extraction resize image by 300 x 350 pixel}
	\label{fig:3pics}
\end{figure}

 \begin{figure}[!h]
\centering
\begin{subfigure}[b]{0.2\linewidth}
 	\includegraphics[width=\linewidth]{origin.png}
	\caption{Original image}
\end{subfigure}
\hspace{5mm}
\begin{subfigure}[b]{0.2\linewidth}
	\includegraphics[width=\linewidth]{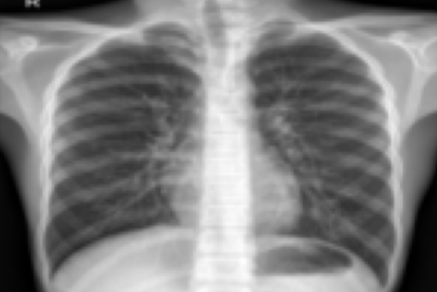}
	\caption{Resize image}
\end{subfigure}
\hspace{5mm}
\begin{subfigure}[b]{0.2\linewidth}
	\includegraphics[width=\linewidth]{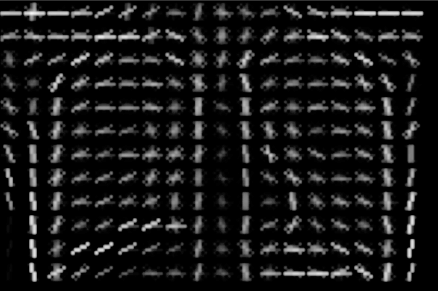}
	\caption{HOG}
	\end{subfigure}
	\caption{HOG feature extraction resize 100 x 150 pixel}
	\label{fig:3pics}
\end{figure}

Figure 5 illustrated in the scanning X- Ray image another technique to highlighted the object in the X-Ray image. By highlighting the object in the X-Ray image easy to see some point in area on object and with kind of transparency can find the changes.
Figure 6 illustrated another challenging technique to detect Region of Interest (ROI) extraction for the HOG. Define the left and right side of the organ by separately shapes. We can select only one side to interest of the object. Figure 6 demonstrated the result.

 \begin{figure}[!h]
\centering
\begin{subfigure}[b]{0.2\linewidth}
	\includegraphics[width=\linewidth]{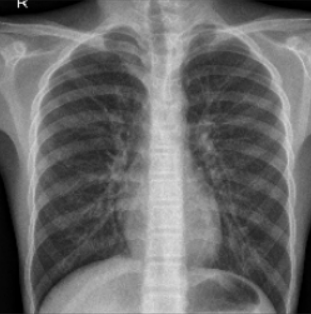}
	\caption{Original image}
\end{subfigure}
\hspace{5mm}
\begin{subfigure}[b]{0.2\linewidth}
	\includegraphics[width=\linewidth]{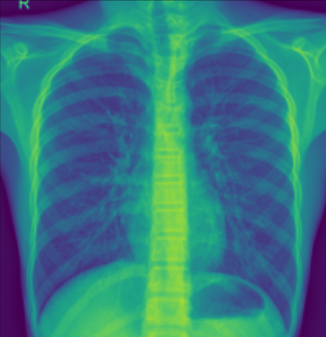}
	\caption{HOG}
	\end{subfigure}
	\caption{Region of Interest (ROI) based on HOG}
	\label{fig:2pics}
\end{figure}

\begin{figure}[!h]
\centering
\begin{subfigure}[b]{0.2\linewidth}
	\includegraphics[width=\linewidth]{orig_C.png}
	\caption{Original image}
\end{subfigure}
\hspace{5mm}
\begin{subfigure}[b]{0.2\linewidth}
	\includegraphics[width=\linewidth]{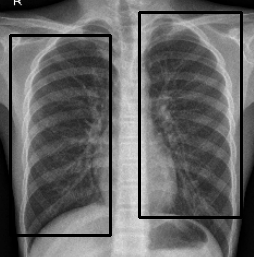}
	\caption{HOG}
	\end{subfigure}
	\caption{Region of Interest (ROI) based on HOG}
	\label{fig:2pics}
\end{figure}



\vfill

 \subsection*{Transfer learning process}
 The transfer learning during the image processing, Convolutional Neural Network (CNN) is considered a challenging process. In the train subset the Artificial Neural Network (ANN) goes through several stages of self-learning, the ANN learns from previous knowledge of data and transform image data information from one neuron to the next neuron.  The input of the neuron is the output of the previous neuron. This briefly explanation how information passes between neurons in the ANN system. Such process is similar to the information translate in the human brain Nero-network system work. The transfer learning process in the image process implies to scanning the image by pixels  starting from upper left side to the right with direction up to down. After several stages of learning the ANN becomes faster and more efficient. This allows to reduce the computational time and errors which allows better performs of the model.
 Using the Adam optimizer, we improved the feature's ability in network learning from the previous layers' and solve optimization problem in the similar problem in the deep learning framework.






\subsection*{RepVGG architecture model transformation from multi-branches to compact and plane model}
In this section, we describe the architectural structure of the Convolutional Neural Network (CNN) named RepVGG. 
Based on the RepVGG model, we create an algorithm which classify the screening X-Ray image in three categories:  Covid-19, Pneumonia and Healthy cases.  and comparison with the multi branches family . 

We also illustrated the transformation the model from multi-branches to plane model. Such transformation and combination of the architecture structure we used to apply to our deep learning models, as a result, show great potential in detecting diseases based on chest X-Ray images. We compare the deep learning methods based on CNN. We present a high-accuracy classification of medical x-ray images and analysed the results.



\subsubsection*{RepVGG}

This work considers a new algorithm that classifies X-Ray images based on the new architecture structure of the Convolutional neural network known as RepVGG. The first time this RepVGG architecture was presented was in January 2021 by Ding et al.\cite{Ding}. The peculiarity of the architecture is that it initially belongs to a branch family, but after the training process it transforms in the size and takes the form of a stack without branches. This new innovative transformation facilitates rapid classification of the new data in the test sub-data set. The basis of the architecture of RepVGG combines the deep Residual model as a branch structure and deep VGG-16 plane structure\cite{ha}\cite{res}. 
We briefly explain the VGG-16 and ResNet structures.
The VGG-16 model is a deep Convolutional neural network that stands for the Visual Geometry Group from Oxford and was proposed by Simonyan et al.in 2014\cite{Simon}. This model is popular in the deep learning community and used for image classification problems. It has 21 total layers but only 16 are activated layers for the training set. The pre-trained model can classify images into categories, such as Pneumonia, COVID-19, and Normal. We teach the algorithm a new model to recognize and classify the images and train the model used for prediction. The data goes through a stack of five blocks of 13 convolutional layers with a max pooling after each block. The last tree fully connected dense layers used the softmax function to perform classification. The computational time is longer since the model goes through all layers. 

The ResNet50 is a deep convolutional neural network described in the paper of He et al. \cite{Res}. The number 50 represents the number of layers in the model.
The  model applied a strategy shortcut by skipping some residual layers.  The transfer process data in the network is much faster but following the analysis shortened path of skipping residual layers leads to weakening of the power and reduced capacity of the learning process  training data set of the network and loss some information of data. The decreasing capacity of the learning model leads to  lower accuracy in the final model.

The Residual model, in general, is much faster than the VGG model, but RepVGG runs the process much faster and more efficiently than ResNet50. 
The combination of these two powerful structures makes the learning process in image processing more significant.

The architecture structure of RepVGG contains five $3 \times  3$ convolutional blocks and 1x1 branches, and ReLU activation function in the training process is shown in Figure 6. ReLU (Rectified Linear Unit) is very often used in deep learning to activate functions in Artificial Neural Networks. The function is linear and increases activated when returning a positive value bigger than zero. Then function activated the information between the current neuron and the next neuron goes through. Figure 8 shows the activation function.

\begin{figure}[!h]
\center
	\includegraphics[scale=0.3]{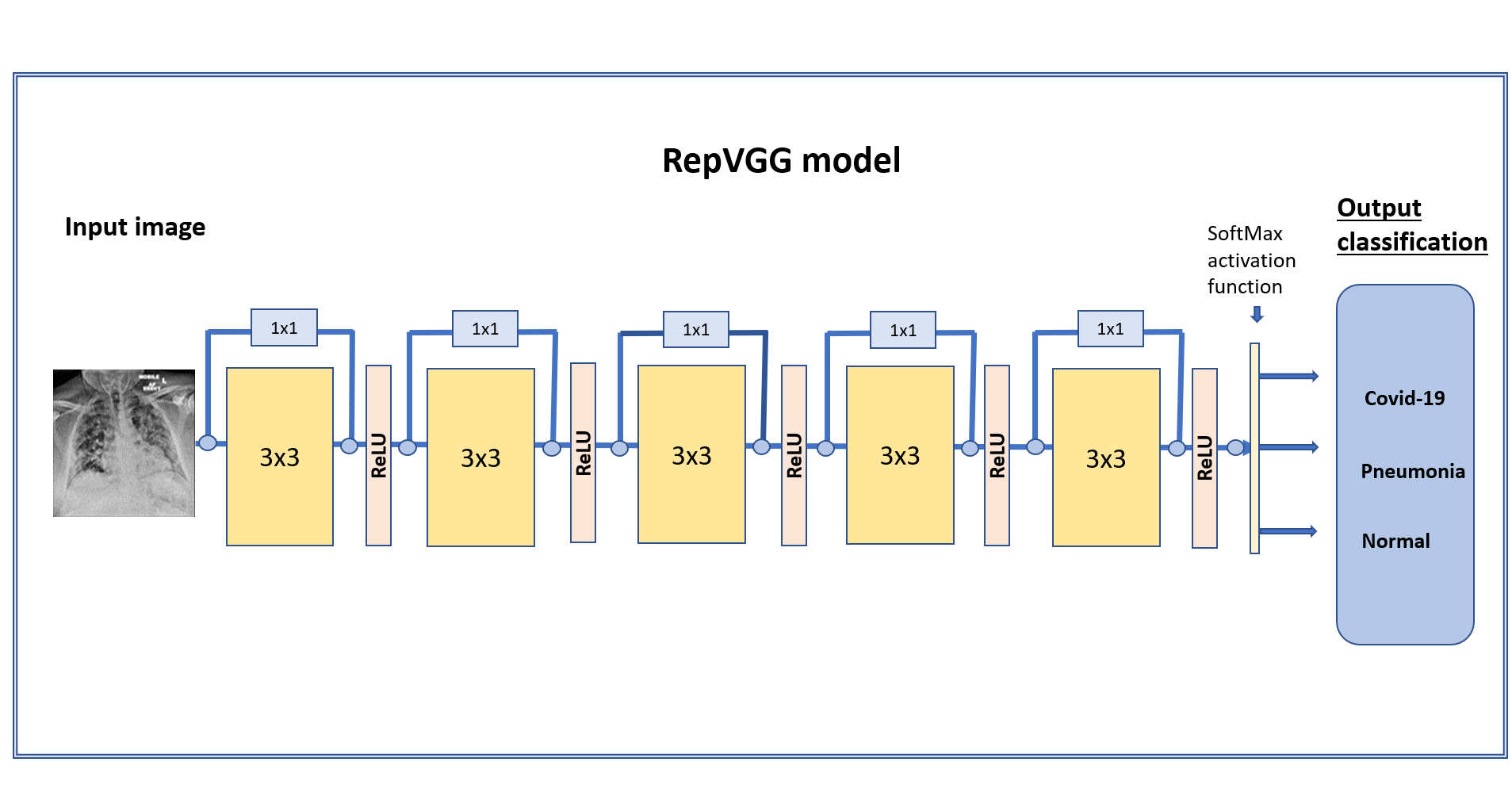}
	\caption{RepVGG architecture model for training}
\end{figure}


The model transforms information in an artificial neural network between neurons during the training data in the learning process. After the training process, the model's architecture changed from multi-branches to a plane model by removing 1x1 branches. 
Figure 7 shows the model after transformation.
\begin{figure}[!h]
\center
	\includegraphics[scale=0.3]{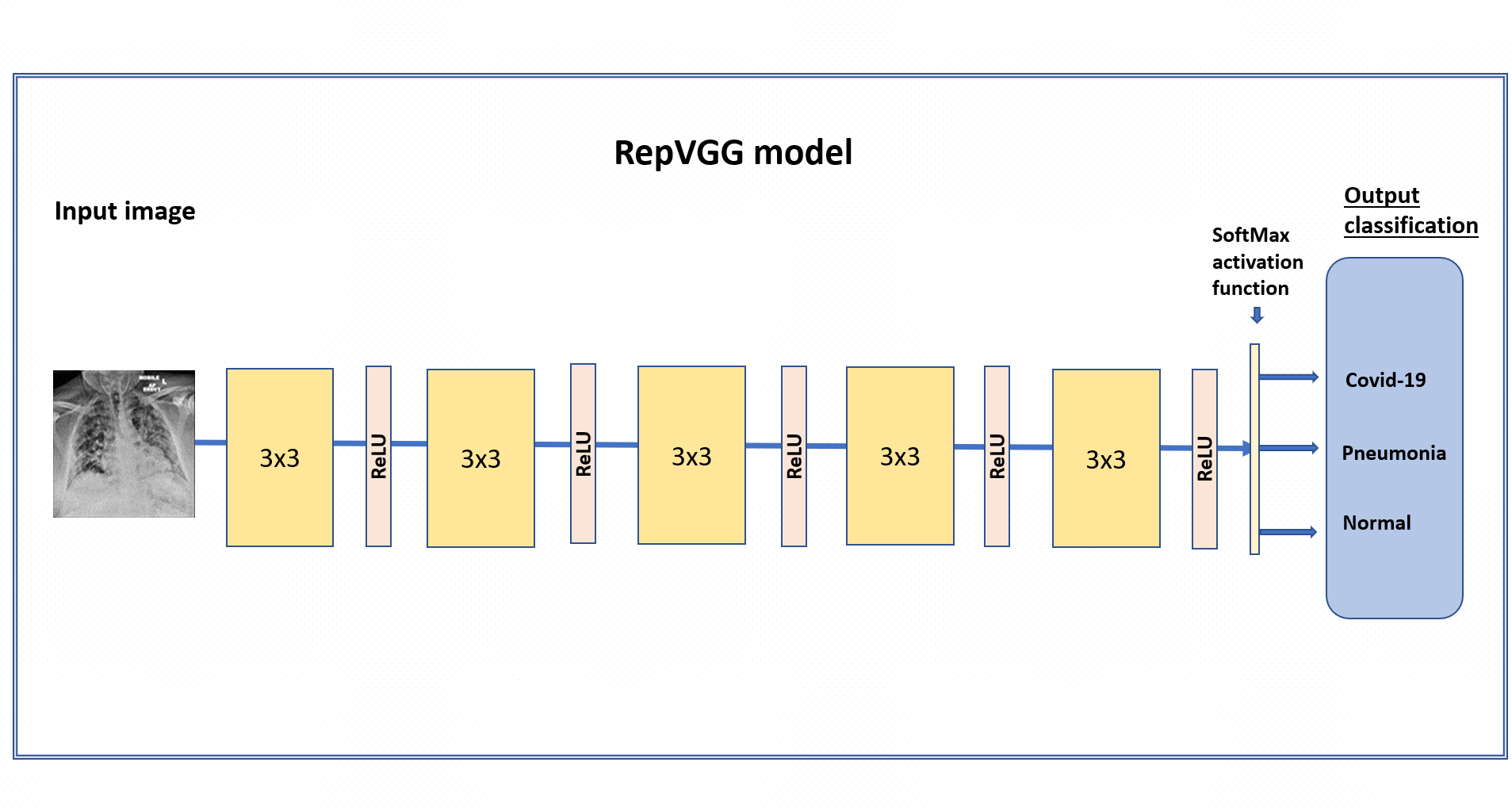}
	\caption{RepVGG architecture model for test}
\end{figure}

We applied and fine-tuned the RepVGG model for COVID-19, Pneumonia, and Healthy cases diagnosis in the X-ray images.

The model transforms information in an artificial neural network between neurons during the training data in the learning process. After the training process, the model's architecture changed from multi-branches to a plane model by removing 1x1 branches. 
This transformation and reducing the size of the RepVGG architecture make this method more efficient and faster.


Because the architecture structure of RepVGG, becomes smaller, the signals of information in the RepVGG architecture are transmitted faster after the training process,  Such transformation of the model saves time and reduces the cost for manufacturers and developers.

\subsection*{Grad-Cam - computer vision visualization technique}

GRAD - CAM Gradient-weighted Class Activation Mapping stands from the uses the calculation of values of gradients and activation function in the last deep learning layer, highlighting the image's localization target regions\cite{grad}.

Applying this technique, we created the Feature map from the training model for each different X-ray image. The Grad-Cam method highlights the affected area of the lungs in COVID-19 or Pneumonia cases and detects the lung in Healthy cases.  We superimposed this created Feature map on the original chest X-ray image and predicted the lung's affected area by visual explanation in the X-Ray image. This Visual explanation can be seen from the network's last convolutional layer, which produces a localization map and highlights the lung's affected area. The last convolutional layer's output shows the disease's prediction in the X-ray images. Artificial Intelligence develops such automatic diagnoses in X-ray images. We improved the detection of specific infection areas or diseases by comparing two deep learning frameworks, RepVGG and ResNet50 models. From the picture, we can see that RepVGG and ResNet50 work well to predict diseases. The RepVGG method illustrated better prediction of COVID-19, pneumonia, and lung of Healthy patients than ResNet50. RepVGG, with the Grad-Cam process, created a Feature map more accurately and shows a much more extensive detection area in X-ray images.  Figures 9 and 10, demonstrated created Feature map and prediction on the X-ray image with the Covid-19 case. The first picture (a), is the original X-ray image from the training model set. In the second image (b), we show the hidden process of creating a Feature map. This process was carried out by calculating the affected area's gradients and highlighting the infection. The third picture (c), demonstrated the final result of the prediction daises in X-ray images using the Artificial Intelligence technique, and this result we will see on the screen.  We compare the Feature map in the X-ray images with two deep learning methods, RepVGG and ResNet50. 
 


\begin{figure}[!h]
\center
	\includegraphics[scale=0.5]{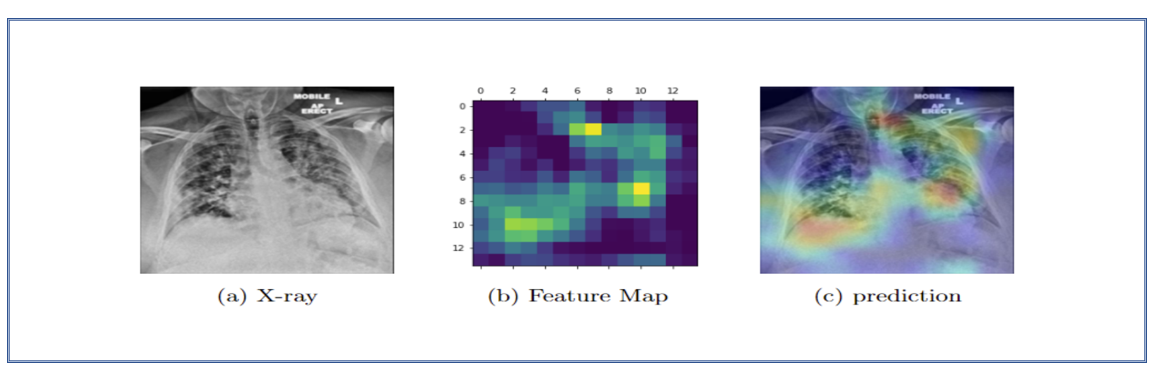}
	\caption{Grad-Cam using RepVGG for Covid-19 case}
\end{figure}

\begin{figure}[!h]
\center
	\includegraphics[scale=0.3]{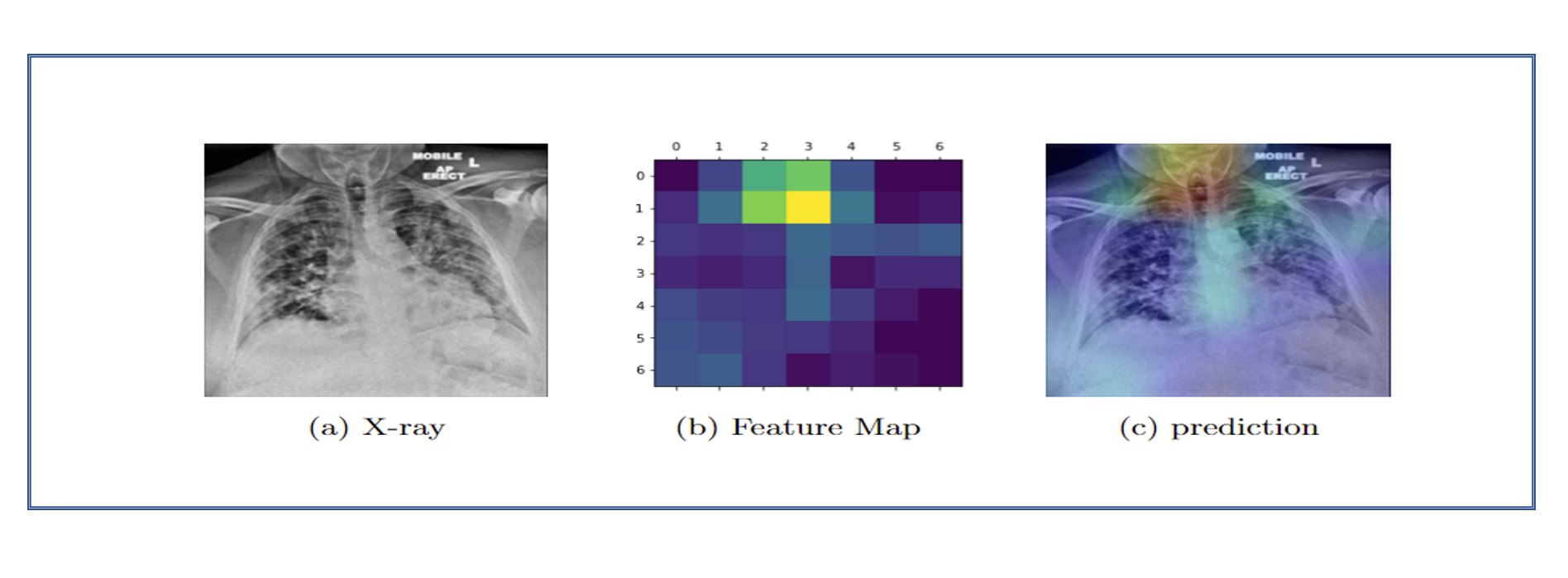}
	\caption{Grad-Cam using ResNet50 for Covid-19 case.}
\end{figure}

\begin{figure}[!h]
\center
	\includegraphics[scale=0.3]{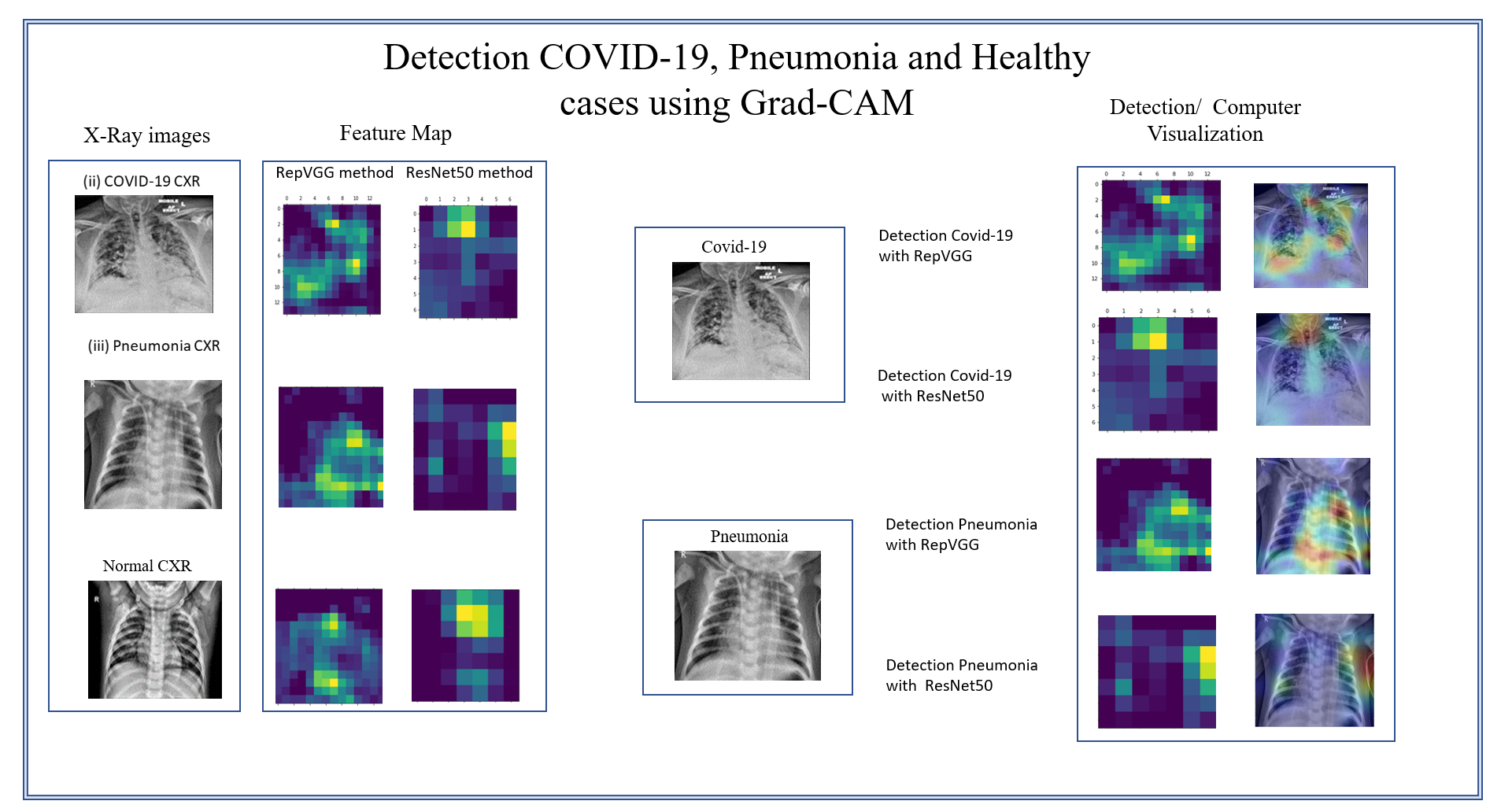}
	\caption{Detection Covid-19 using AI and computer vision.}
\end{figure}

\begin{figure}[!h]
\center
	\includegraphics[scale=0.3]{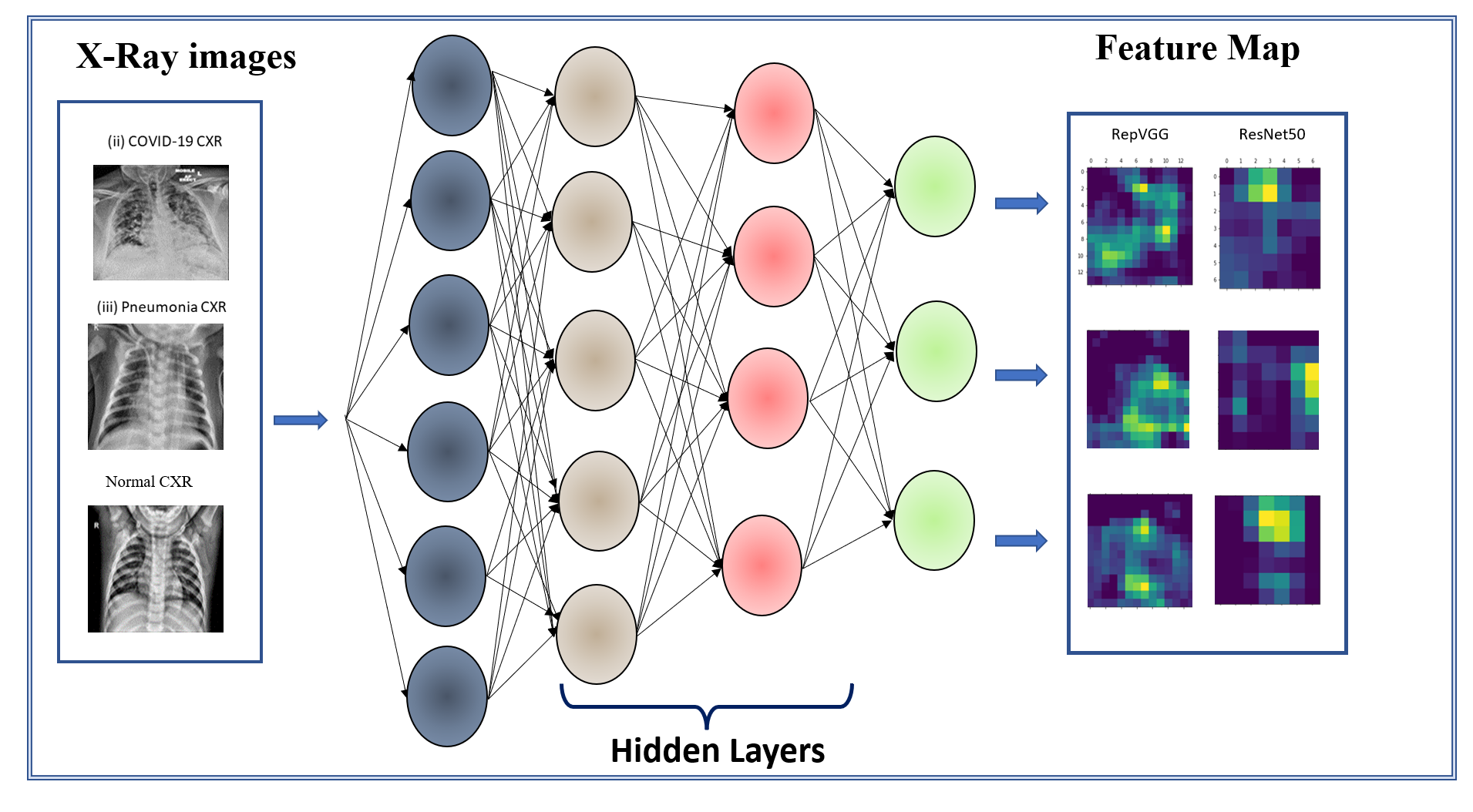}
	\caption{Deep learning for image processing.}
\end{figure}

\section{Results and discussions.}

In this section, we discuss and describe the concept of the prediction results of the classification of x-ray images. As we show in During the training process, the training model classifies the x-ray images into tree categories artificial neural network model. Then we use this model for the testing data set. To investigate the results of the accuracy of the training model, we represent the result in several ways. In particular, we introduce the concept of Confusion matrix, Classification report, ROC -AUC evaluation by tables, graphs, and matrix.




We created a confusion matrix table to visualize and estimate the classification machine learning model for the multi-class classification. This matrix represents numerical values prediction  correct and wrong classification x-ray images. This task is to classify X-ray images into three categories: Covid-19, Pneumonia, and Healthy cases. We visualize the classification process as output three rows by three columns matrix and summarize and explain the result in numerical value. The rows demonstrated the Actual true label of images, and the columns demonstrated the Prediction label after the classification process. The diagonal of the matrix shows the correct prediction, i.e. predicted Positive for sick people (true positive) in numerical values. The numbers off-diagonal show the model’s wrong predictions such as healthy people predicted positive (false positive) in numerical value. The accuracy of the model summarize in the matrix and demonstrated in four metrics:
\begin{enumerate}
    \item TP (true positive)  - the model correctly predicts the diseases.
    \item TN (true negative)  - the model's correct prediction of the healthy cases true negative.
    \item FP (false positive) - the model wrong predicts the positive cases. Then the X-ray image of                                   healthy cases predicts diseases.
    \item FN (false negative) - the model is wrong to predict the diseases, then the X-ray image mark actually                               the true label of the disease, but the model can't predict it.  
\end{enumerate}
The summary prediction of the model is in the following Figure 12. The accuracy of our model can be calculated with the formula: number of correct predictions by total number of predictions. The accuracy is the


$$\text{Accuracy} =  \frac{TP+TN}{TP+TN+FP+FN}.$$

$$\text{F1 - score} =  \frac{2TP}{2TP+FP+FN}.$$

$$\text{Precision} =  \frac{TP}{TP+FP}.$$

$$\text{Sensitivity} =  \frac{TP}{TP+FN}.$$
\begin{figure}[!h]
	\center
	\includegraphics[scale=0.45]{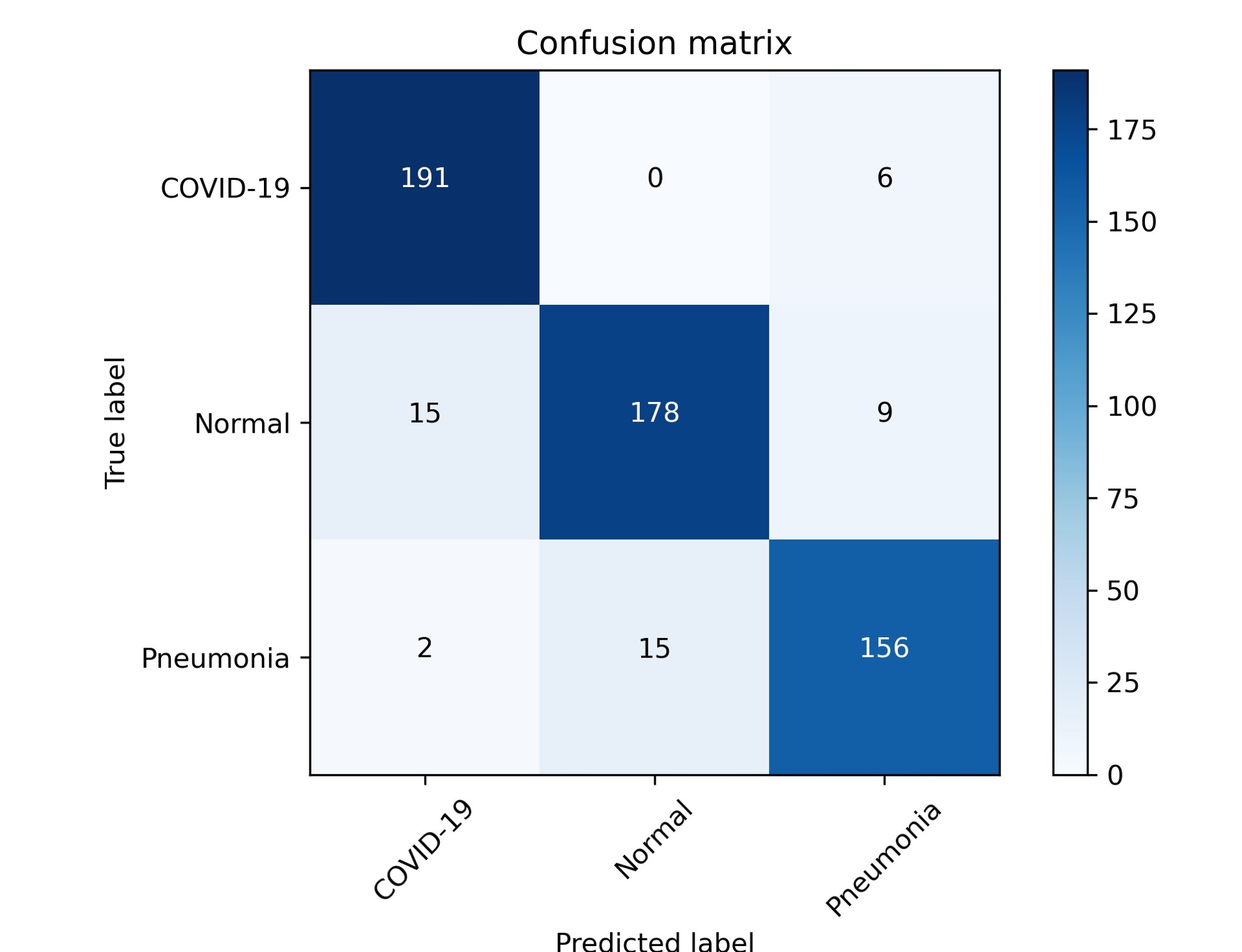}

		\caption{Confusion matrix.}
 \end{figure}


\subsection*{AUC - ROC evaluation, accuracy and loss validation of the training data set.}
In this section, we discuss the evaluation features. For the accuracy of the training model, we measured with another popular machine learning technique as AUC - ROC. 
To demonstrate the model and how it works we evaluate the data by Receiver operating characteristic (ROC) curve evaluation. We demonstrated the result of accuracy by graphing and visualizing how the model is work. We represented accuracy for three categories: Covid-19, Pneumonia, and Healthy cases. Figure 12 illustrated the prediction of the model with the high accuracy detection and classification Covid-19 classifies correctly with 99\%,  Normal cases classify with an accuracy of 97\%, and Pneumonia also classifies well with 95\%  X-Ray images. Every number above 95\% is considered a good classification. The curve close to 1 demonstrated the best accuracy, and in our work accuracy is very close to 1 which means close to 100\%. 

The results we obtained from RepVGG  provided in the Result section and Tables 1, 2, and 3 in Figures 14, 15, 16 and show better accuracy classification in X-Ray images than other models/ And, through the comparison with popular deep learning models, i.e., VGG, ResNet 50, inceptionV3, DenseNet, and Inception ResnetV2, the proposed framework shows the best diagnostic accuracy\cite{deep}\cite{rink}\cite{goin} 

\begin{figure}[!h]
\centering
\begin{subfigure}[b]{0.27\linewidth}
 	\includegraphics[width=\linewidth]{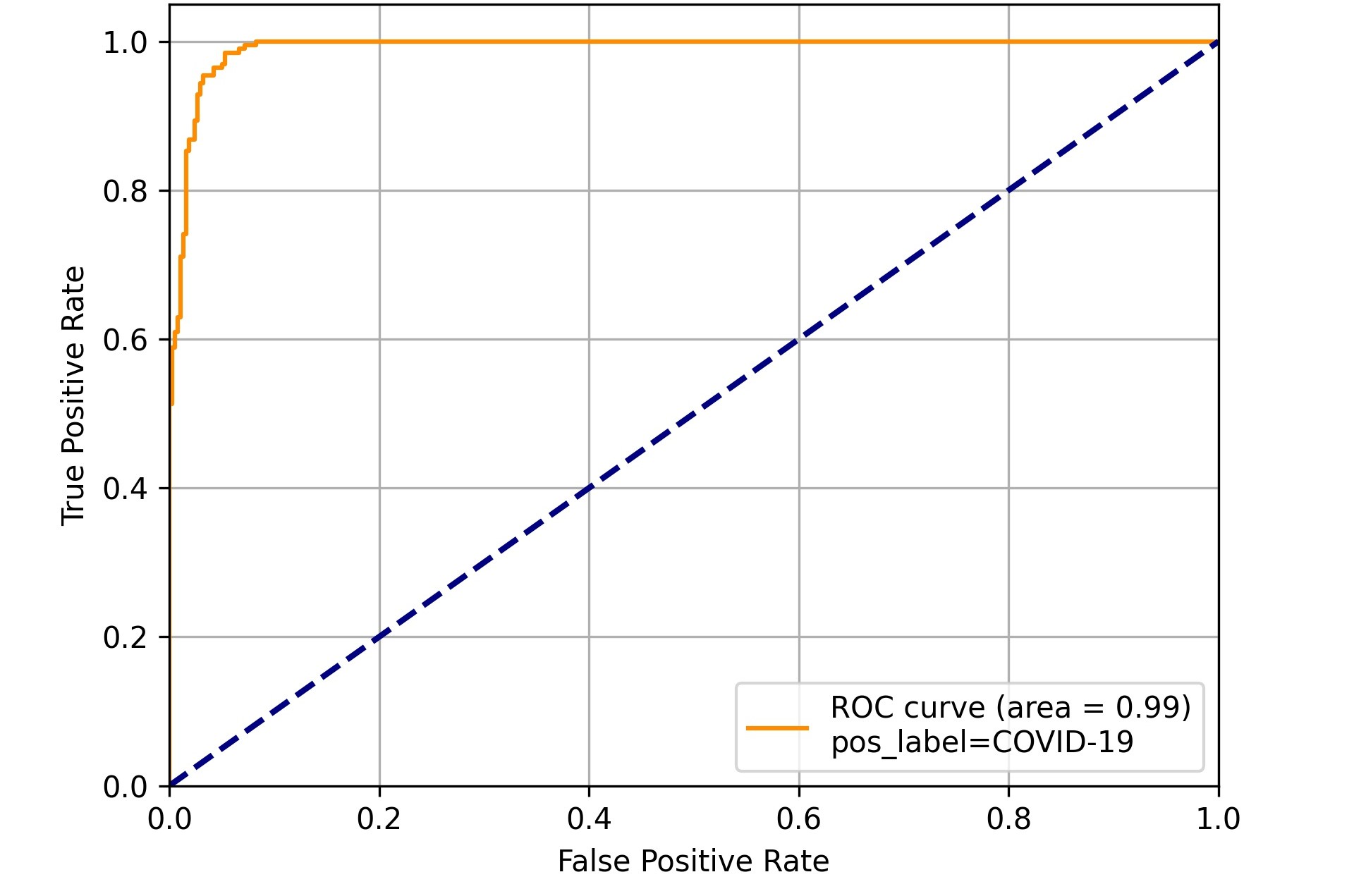}
	\caption{Prediction accuracy of Covid-19 cases}
\end{subfigure}
\hspace{5mm}
\begin{subfigure}[b]{0.3\linewidth}
	\includegraphics[width=\linewidth]{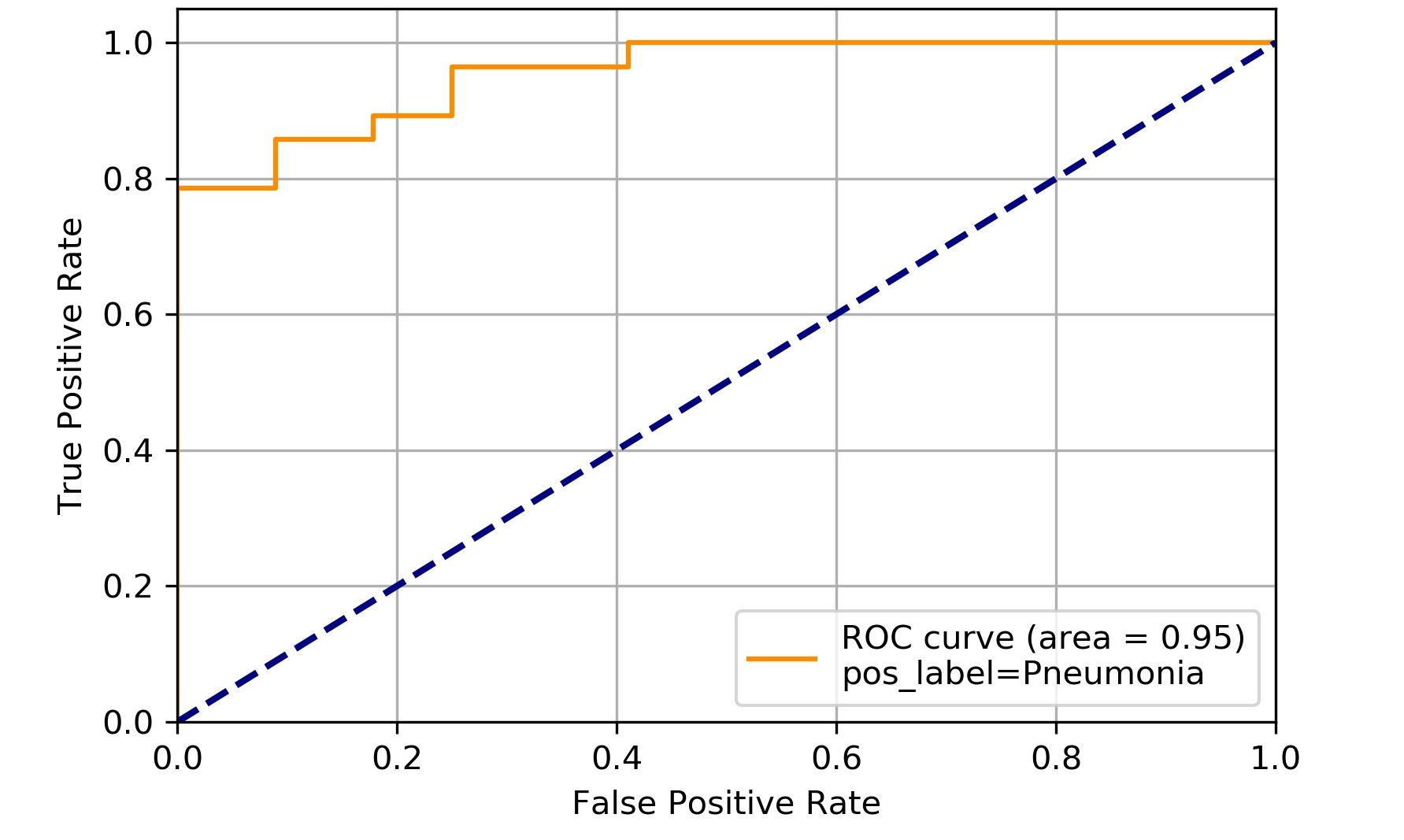}
	\caption{Prediction accuracy of pneumonia cases}
\end{subfigure}
\hspace{5mm}
\begin{subfigure}[b]{0.3\linewidth}
	\includegraphics[width=\linewidth]{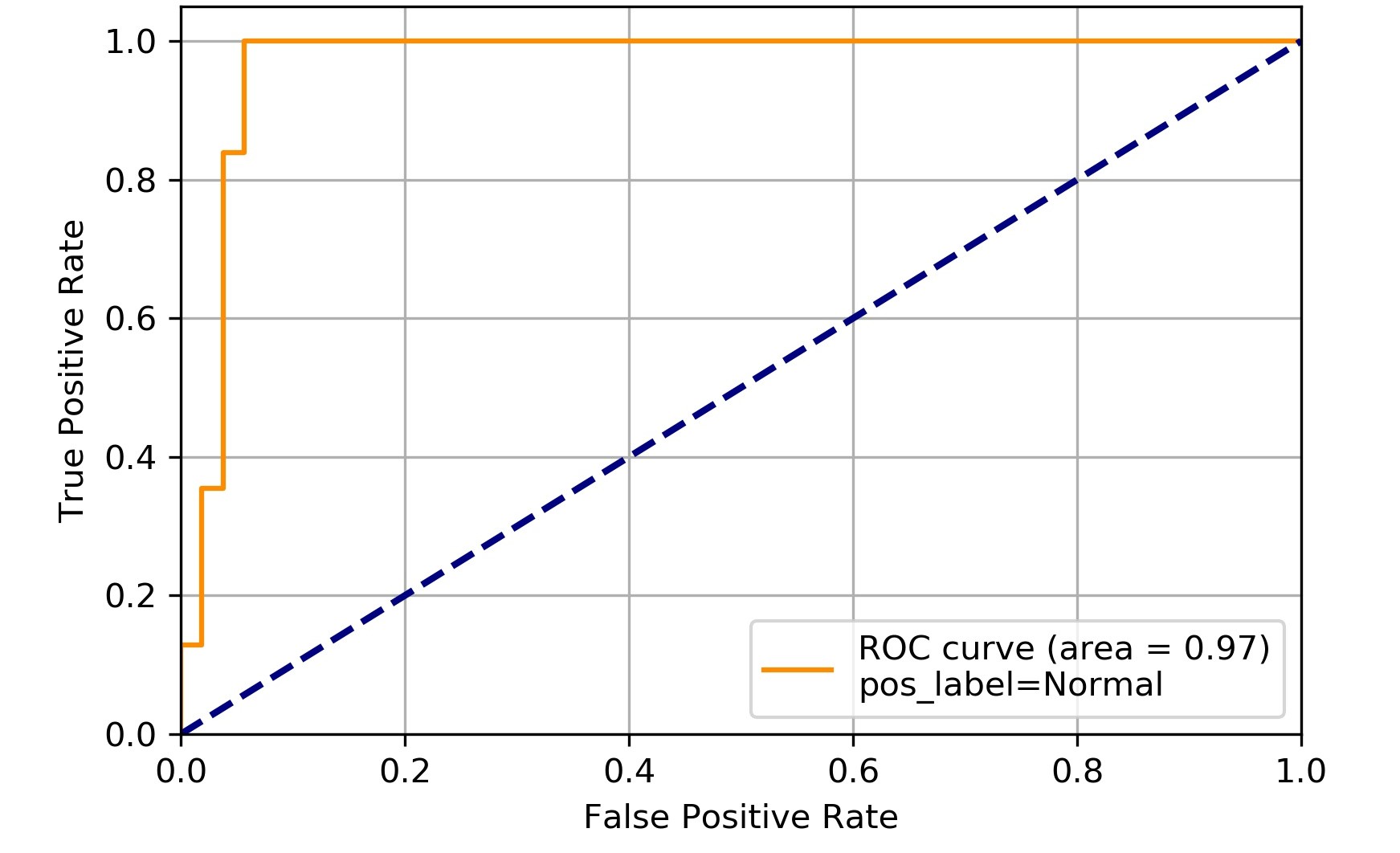}
	\caption{Prediction accuracy of healthy cases}
	\end{subfigure}
	\caption{ The Roc curve accuracy prediction for the three cases Covid-19, Pneumonia, and Healthy patients}
	\label{fig:3pics}
\end{figure}

\subsection*{Model accuracy and loss validation of the training data set.}
During the training process, the data was randomly split with a small ratio of 0.2 from the validation sub-data training set. We trained the validation subset with 20 epochs and predict the accuracy of the model. Figure 13 (a) shows the loss and validation data during the training process. The validation model shows decreasing error and increasing accuracy.  The middle figure (b) shows increasing accuracy during the training process. In the beginning, the blue line illustrated the lower accuracy of the training model, but with each interaction, the model increased the better accuracy. Figure 13 (c) illustrated the same training set but with 50 epochs. In these cases, with 20 and 50 epochs the model shows high numbers of correct classification X-ray images. 

\begin{figure}[!h]
\centering
\begin{subfigure}[b]{0.3\linewidth}
 	\includegraphics[width=\linewidth]{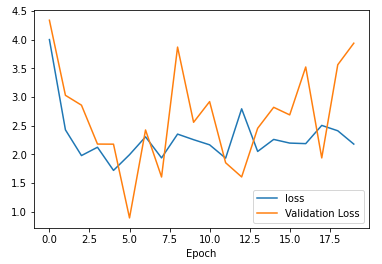}
 	\caption{loss and validation loss}
\end{subfigure}
\hspace{5mm}
\begin{subfigure}[b]{0.3\linewidth}
	\includegraphics[width=\linewidth]{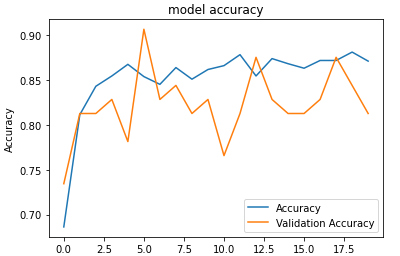}
	\caption{Model accuracy of the RepVGG training model}
\end{subfigure}
\hspace{5mm}
\begin{subfigure}[b]{0.3\linewidth}
	\includegraphics[width=\linewidth]{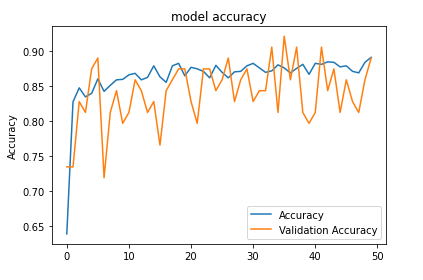}
	\caption{Models accuracy for the training model with 50 epoch}
	\end{subfigure}
	\caption{Loss and validation and accuracy in training model for the three cases Covid-19, Pneumonia, and Healthy patients}
	\label{fig:3pics}
\end{figure}

\subsection*{Table of comparison performing classification x-ray images}
The deep learning RepVGG model shows great potential in detecting COVID-19 based on chest X-ray images. The comparison with popular deep learning models, such as VGG, ResNet 50, inceptionV3, DenseNet, and Inception ResnetV2, the proposed framework of the RepVGG model, shows the best diagnostic accuracy. From the tables in Figure 14, Figure 15, and Figure 16, we can note that the deep learning RepVGG method shows the best X-ray classification result in three cases. The accuracy for classification Covid-19 is 0.9579\% is much higher in comparison Inception ResnetV2 0.9334\% , DenseNet 0.9281\%, VGG16 0.9089\%, InseptionV3 0.8844\% and ResNet50 0.8563\%. Table Figure 15, classification x-ray images with pneumonia diseases RepVGG also show height accuracy of 0.9579\%. Figure 16 illustrates the same great high-accuracy classification of the Normal cases.
\begin{figure}[!h]
\center
	\includegraphics[scale=0.5]{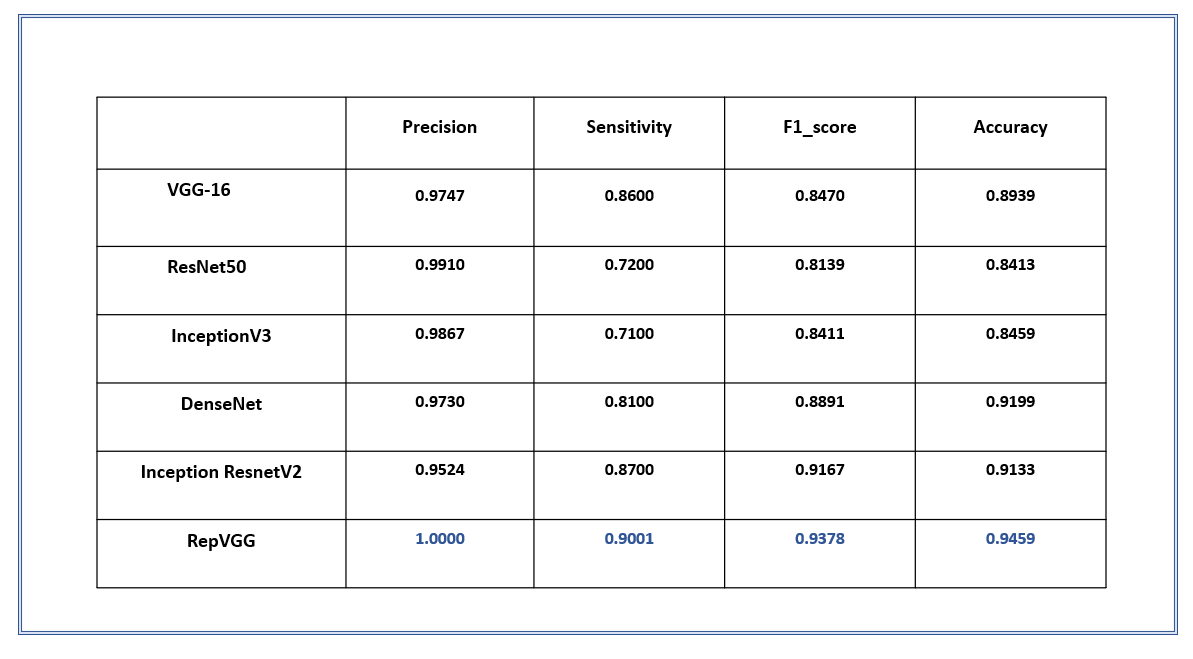}
	\caption{Table of comparison deep learning methods for classifying Covid-19 X-Ray images.}
	
\end{figure}


\begin{figure}[!h]
\center
	\includegraphics[scale=0.5]{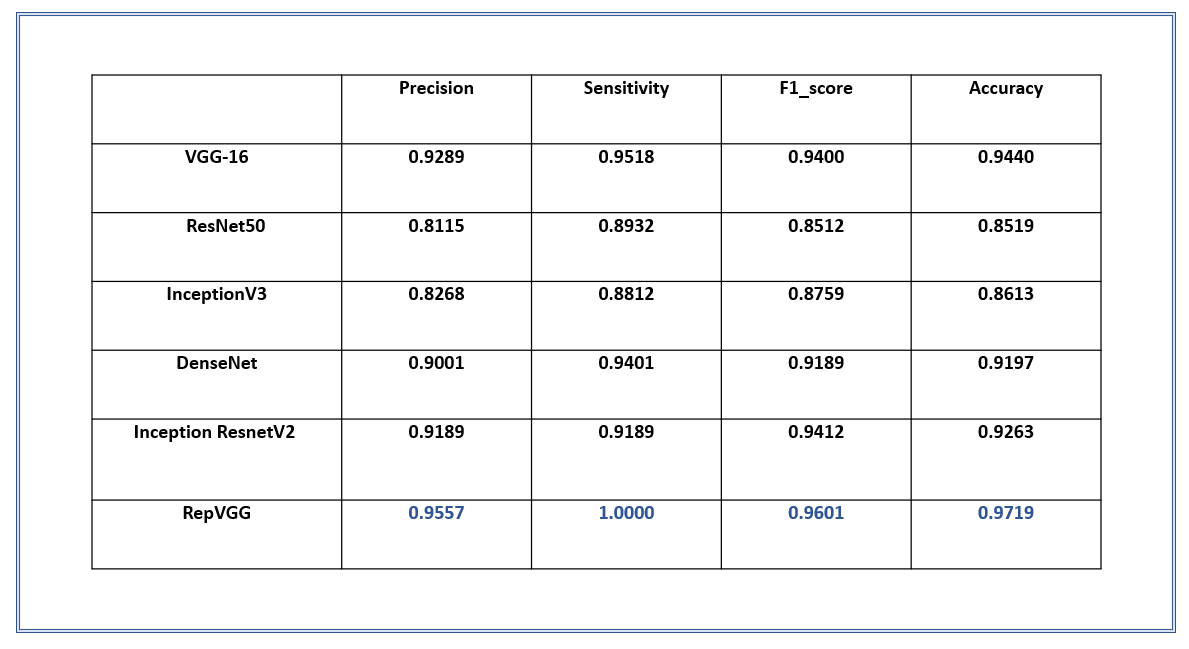}
	\caption{Table of comparison deep learning methods for classifying Pneumonia X-Ray images.}
\end{figure}


\begin{figure}[!h]
\center
	\includegraphics[scale=0.5]{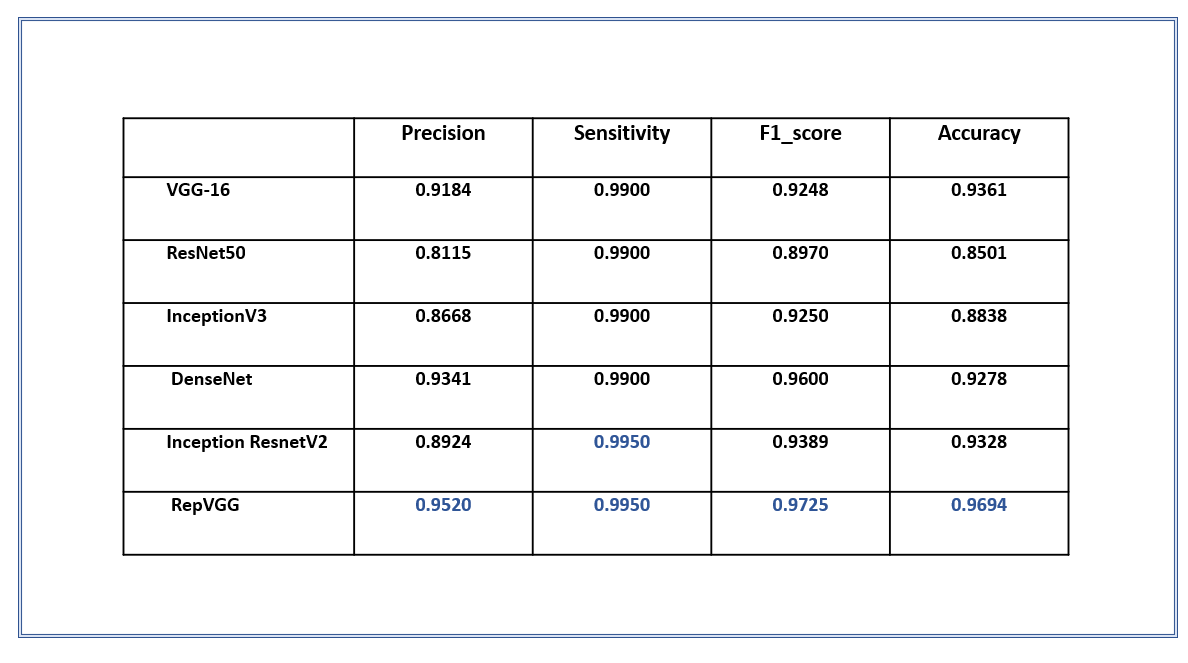}
	\caption{Table of comparison deep learning methods for classifying Normal X-Ray images.}
\end{figure}

\section{Conclusion.}
Applying AI and Machine Learning techniques in the medical field could benefit from rapidly performing challenging tasks. The proposed framework can accurately classify X-ray images as an example of COVID -19, Normal, and Pneumonia. The classification accuracy of the proposed model achieves 95.79\% for three classes. With a lung region detection and RepVGG-based deep learning framework, the system can accurately identify lung regions and discriminate lung infection categories in X-Ray images. This approach could help a vast medical lot classify and detect diseases such as cancer, skin, cancer breast, or abnormal processes in the image tasks.
	
		
		
		
		
		



\subsection*{Acknowledgement}

It is the pleasure to thank the Center of Undergraduate Excellence (CUE) at Chapman University for the Summer Undergraduate Research Fellowship (SURF) grant which supported part of this research. It is also a pleasure to thank Dr. Yuxin Wen Assistant Professor, Fowler School of Engineering at Chapman University for her guidance and advising during the writing of this paper, and Dr.Erik Linstead Associate Professor, Senior Associate Dean Fowler School of Engineering; Electrical Engineering and Computer Science for his help and support.


\end{document}